# An Automatic Detection Method Of Cerebral Aneurysms In Time-Of-Flight Magnetic Resonance Angiography Images Based On Attention 3D U-Net

Geng Chen[a,b,*], Chen Meng[d], Di Ruoyu[c], Wang Dongdong[c], Yang Liqin[c], Xia Wei[a,b], Li Yuxin[c,#], Geng Daoying[c,#]

[a] *Academy for Engineering and Technology, Fudan University, 20 Handan Road, Shanghai, 200433, China*

[b] *Suzhou Institute of Biomedical Engineering and Technology, Chinese Academy of Sciences, 88 Keling Road, Suzhou, 215163, China*

[c] *Department of Radiology, Huashan Hospital, Fudan University, 12 Wulumuqi Middle Road, Shanghai, 200040, China*

[d] *Biomedical Engineering, Xuzhou Medical University, 209 Tongshan Road, Xuzhou, 221004, China*

## Abstract

**Background:** Subarachnoid hemorrhage caused by ruptured cerebral aneurysm often leads to fatal consequences. However, if the aneurysm can be found and treated during asymptomatic periods, the probability of rupture can be greatly reduced. At present, time-of-flight magnetic resonance angiography is one of the most commonly used non-invasive screening techniques for cerebral aneurysm, and the application of deep learning technology in aneurysm detection can effectively improve the screening effect of aneurysm. Existing studies have found that three-dimensional features play an important role in aneurysm detection, but they require a large amount of training data and have problems such as a high false positive rate.

**Methods:** This paper proposed a novel method for aneurysm detection. First, a fully automatic cerebral artery segmentation algorithm without training data was used to extract the volume of interest, and then the 3D U-Net was improved by the 3D SENet module to establish an aneurysm detection model. Eventually a set of fully automated, end-to-end aneurysm detection methods have been formed.

**Results:** A total of 231 magnetic resonance angiography image data were used in this study, among which 132 were training sets, 34 were internal test sets and 65 were external test sets. The presented method obtained 97.89±0.88% sensitivity in the five-fold cross-validation and obtained 91.0% sensitivity with 2.48 false positives/case in the detection of the external test sets.

**Conclusions:** Compared with the results of our previous studies and other studies, the method in this paper achieves a very competitive sensitivity with less training data and maintains a low false positive rate. As the only method currently using 3D U-Net for aneurysm detection, it proves the feasibility and superior performance of this network in aneurysm detection, and also explores the potential of the channel attention mechanism in this task.

## 1. Introduction

Subarachnoid hemorrhage(SAH) caused by cerebral aneurysm often causes death or severe disability. Although it can cause such serious consequences, cerebral aneurysms do not appear suddenly. They usually have an incubation period of several years or even decades, during which they do not show any symptoms.[1] Treatments such as clipping and endovascular intervention when the cerebral aneurysm is not ruptured can get a better prognosis and significantly prolong the patient's survival. Therefore, regular aneurysm screening is performed and cerebral arteries are found in the asymptomatic stage intervention or treatment of aneurysms in time is one of the effective ways to avoid aneurysm rupture. Time-Of-Flight magnetic resonance angiography(TOF-MRA) is currently one of the most commonly used methods for screening aneurysms. Because it has diagnostic accuracy similar to DSA and CTA[2, 3], and is a non-invasive examination method, it is especially suitable for aneurysm screening when asymptomatic.

For radiologists, it has always been an arduous task to screen aneurysms quickly, massively, and accurately in scenarios such as physical examinations. Computer Assisted Detection methods provide a promising solution for aneurysm screening.[4-12] In this field, several works have been proposed in recent years. Joseph Stember et al.[11] detected 98.8%(85/86) basilar tip aneurysms in 1.5T and 3.0T TOF-MRA images. Faron Anton et al.[12] got 90% sensitivity at 6.1 False Positive(FPs)/case. Nakao Takahiro et al.[8] detected 94.2% (98/104) of aneurysms with 2.90 FPs/case, with sensitivity of 70.0% at 0.26 FPs/case. Ueda Daiju et al.[9] got 91% sensitivity at 6.60 FPs/case. In the above studies, researchers have found that three-dimensional features have an important impact on the performance of aneurysm detection methods. Faron et al.[12] used the Deepmedic framework[13] with three-dimensional convolution. Joseph et al.[11] and Nakao et al.[8] used a 2D CNN network, but their inputs were multi-angle slice sets for each sample block. The above works prove that the three-dimensional convolutional neural network is suitable for the task of cerebral aneurysm detection. However, the main disadvantage of 3D convolutional networks in application is that it is difficult to fully optimize the network hyperparameters with a small training data set. And most of the time, collecting a large amount of TOF-MRA image data and labeling it is a task that requires a lot of manpower. Besides, 3D convolutional networks will use more computing resources, often redundant, and more model parameters, which is not conducive to solving medical imaging related tasks. Attention mechanism was proposed to solve such problems. Among them, SKNet[14], SENet[15], and GCNet[16] are three representative attention models.

In the research of our last paper[17], we proposed an automated computer assisted detection system for cerebral aneurysms using an improved 3D U-Net[18-20]. In which, the method achieved 82.9% sensitivity at 0.86 FPs/case, with 76 cases as training dataset. To improve the performance of the method, in this paper, we compared the effects of adding different attention modules to the original network, and further optimized the best solution. At the same time, we also increased the amount of data in the dataset to improve the performance of the method.

## 2. Methods

### 2.1 Materials

The ethics board of our institution comprehensively re-viewed and approved the protocol of this study. Two of the authors of this paper (D.R. and W.D.) are radiologists with 4 years of work experience. They annotated all the aneurysms in this study, with the DSA as ground truth.

### 2.2 Datasets

A total of 231 patients (all have un-ruptured cystic aneurysm) underwent contrast unenhanced 3.0T 3D TOF-MRA. In this study, angiography examinations were performed with two 3.0T system, GE Discovery MR750 and SIMENS Verio, using two imaging factors separately. On GE MR750 the factors were: repetition time/echo time, 25msec/5.7msec; flip angle, 20°; field of view, 220mm; section thickness, 1.2mm; acquisition matrix, 320×256, reconstructed to 1024×1024×240; acquisition time, 2min14s; and on SIMENS Verio, the factors were: repetition time/echo time, 22msec/3.6msec; flip angle, 18°; field of view, 240mm; section thickness, 0.5mm; acquisition matrix, 384×254，reconstructed to 768×536×162.

Table 1: Detail imaging factors for 3D TOF-MRA in this study

|  | GE MR750 | Siemens Verio |
|---|---|---|
| **TR/TE** | 25msec/5.7msec | 22msec/3.6msec |
| **Flip Angle** | 20° | 18° |
| **FOV** | 220mm | 240mm |
| **Thickness** | 1.2mm | 0.5mm |
| **Acquisition Matrix** | 320×256 | 384×254 |
| **Acquisition Time** | 2min14s | 2min52s |
| **Reconstructed** | 1024×1024×240 | 768×536×162 |

Patients were selected randomly from outpatient and physical examinations, with a period from 2016.03 to 2019.04. All the sets were annotated by painting in the aneurysm area. Then the patients were divided into three datasets: Training dataset, Internal Test dataset, and External Test dataset. Since the training dataset and internal test dataset were used for 5-fold cross-validation, the two parts of data were combined for statistics. Among the 166 patients in training dataset and internal test dataset, 110 were female and 56 were male, age ranged from 23 to 86. Among these patients, 40.4% were over 60 years old. The max diameter of aneurysms ranged from 1.39mm to 23.96mm, and 41.0% of which were under 5mm. The distribution of aneurysms covered the internal carotid artery area, middle cerebral artery area, anterior cerebral artery area, posterior cerebral artery area, basilar artery area, and vertebral artery area. In the training dataset and the internal test dataset, there were 183 aneurysms (13 patients had double cases, 2 patients had triple cases, and 151 patients had single cases). The aneurysms' average size was 6.96mm in the internal carotid artery area, 6.06mm in the anterior cerebral artery area, 8.22mm in the middle cerebral artery area and 6.91mm in the basilar artery area, 12.78mm in the vertebral artery area, respectively. In the above areas, the largest aneurysm was located in the vertebral artery area.

The external test data set was acquired with the same factors like the training dataset and internal test dataset. Among the patients in the external test sets (65 patients totally, age ranged from 17 to 76), 43 were female (age ranged 17-76 years; mean age, 54±14) and 22 were male (age ranged 42-75 years; mean age, 58±9). Among these patients, 40% were over 60 years old. The max diameter of aneurysms ranged from 1.74mm to 40.00mm, and 43.1% of which were under 5mm. The distribution of aneurysms covered the internal carotid artery area, middle cerebral artery area, anterior cerebral artery area, posterior cerebral artery area, but no basilar artery area and vertebral artery area. The aneurysms' average size was 7.51mm in the internal carotid artery area, 5.49mm in the anterior cerebral artery area, 8.99mm in the middle cerebral artery area and 2.92mm in the posterior cerebral artery area, respectively. In the above areas, the largest aneurysm was located in the internal carotid artery area.

Table 2: Detail characteristics of training dataset+ internal test dataset and External test dataset

| Characteristics | Training Dataset + Internal Test Dataset | External Test Dataset |
|---|---|---|
| No. of patients | 166 | 65 |
| No. of male patients | 56 | 43 |
| No. of female patients | 110 | 22 |
| Mean age(y) | 56±11 | 55±12 |
| Male patients | 56±11 | 58±9 |
| Female patients | 56±11 | 54±14 |
| Hypertension patients | 76 | 30 |
| No. of aneurysms | 183 | 67 |
| Mean size of aneurysms | 7.09±4.56 | 7.15±6.51 |
| Size of aneurysms: | | |
| <3.0 | 20 | 11 |
| 3.0-4.9 | 55 | 16 |
| 5.0-9.9 | 72 | 29 |
| >=10.0 | 36 | 11 |
| Location of aneurysms: | | |
| Internal carotid artery area | 90 | 32 |
| Middle cerebral artery area | 36 | 17 |
| Anterior cerebral artery area | 26 | 13 |
| Posterior cerebral artery area | 19 | 5 |
| Basilar artery area | 9 | 0 |
| Vertebral artery area | 3 | 0 |

In order to perform 5-fold cross-validation, we randomly took 132 cases in the training dataset for training and 34 cases for internal testing in each fold experiment. Then augmented the training dataset with discrete Gaussian noise filter(variance: 4.0, max kernel width: 32 pixels), flipping(by transverse section) and histogram equalization filter in turn, and finally got 1056 image sets for training. Then we resampled all the image sets to isotropic and cropped them so the no content edges would not affect the training.

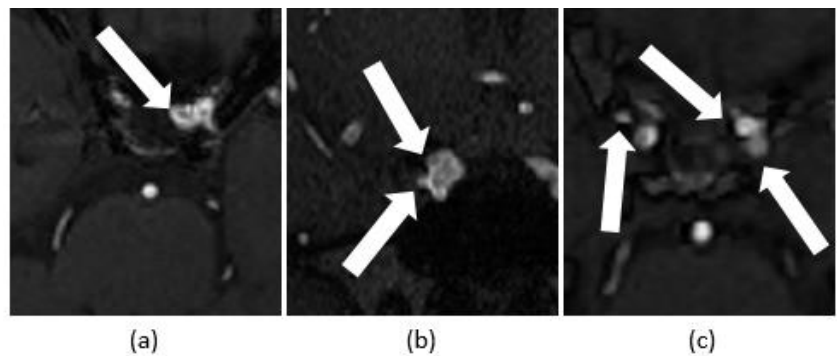

Figure 1: Aneurysms in our Dataset (a) single aneurysms (b) double aneurysms (c) triple aneurysms

**2.3 Development of the method**

In our method, we designed two main steps: First, automatically extraction of the volume of interest(VOI) from the input image; Second, detection of the suspected aneurysm areas using deep neural networks. After completing the training of the deep neural network and obtaining the model, this method realized the automatic detection of cerebral aneurysms.

*2.3.1 Step One: Extraction of the VOI*

The input was DICOM datasets, which in the form of a volume. First, the image grayscale range was normalized to [0, 1024]. Then using a threshold based filter we filtered the image into a binary image(voxels with value 0 and 1). In this study, the threshold of the filter was 300. Took 60% of the slices in the middle of all slices in the Z-axis direction for the next step. For each of the left slices, with the center of the slice as the center, a straight line passing through the center was set at intervals of 30° to obtain 12 straight lines. For each straight line, detected the points on which the voxel density value stepped. For the two furthest points on each line, used them to construct the seed point template. Calculated the distance between the points with the center, if the distance was larger than 25% of the distance from the edge of the image to the center, then the point was selected as one of the boundary points of the seed point area, else created a point at the 25% of the distance as the boundary point. After processing all the left slices in this way, we got a spherical-like region, and the voxels with value 1 in this region were the seed points for region growth. Took the voxel set of the corresponding positions of these seed points on the normalized image, and calculated the average and standard deviation of their intensity value. Since the intensity value distribution of voxels in the blood vessel area in the TOF-MRA image conforms to the superimposed form of multiple Gaussian distributions.[21]

$$f(x) = \frac{1}{\sqrt{2\pi}\sigma} e^{-\frac{(x-\mu)^2}{2\sigma^2}}$$

We assumed that the blood vessel intensity distribution conformed to the Gaussian distribution as above, and used the average and standard deviation obtained by statistics as the basis for calculating the intensity distribution range of the voxel of the blood vessel region in the image. By calculating the area under the curve of the Gaussian distribution, the maximum and minimum voxel density of the blood vessel area was estimated. In our study, took $\mu - 1.28\sigma$ and $\mu + 1.28\sigma$ as the minimum and maximum of the threshold for region grow, $\mu$ represented the average value and $\sigma$ represented the standard deviation. The area under the curve of the Gaussian distribution function of the area between these two values was 80% of the total area. Using the threshold we did the auto-threshold region grow from the seeds, and dilated the result area with a spherical kernel, then the volume of interest was extracted from the normalized image.

9 / 20

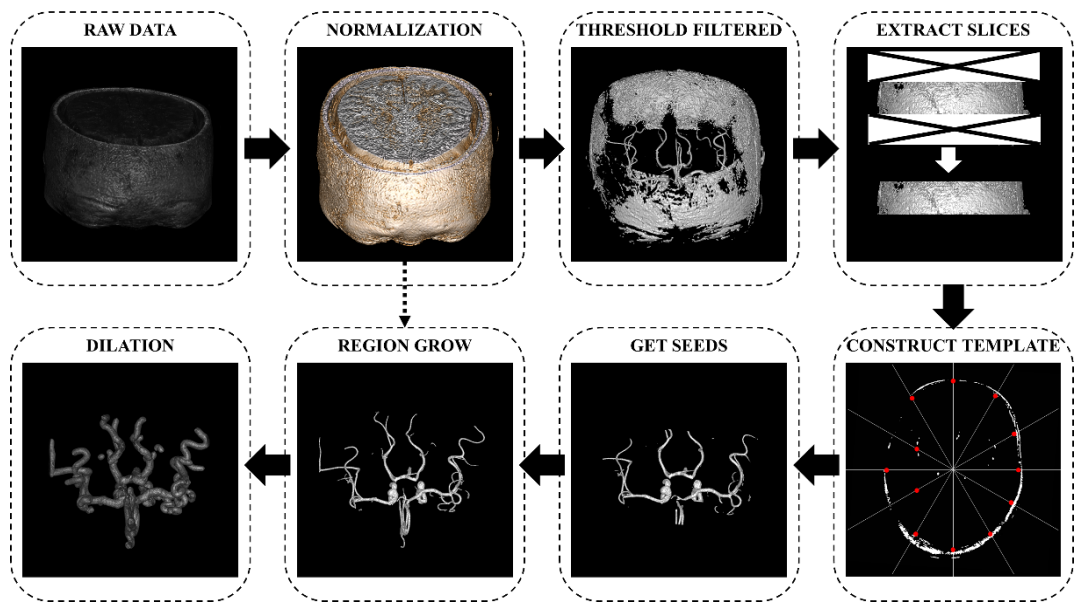

Figure 2: Workflow of the extraction of the VOI

### 2.3.2 Step Two: Detection of the suspected aneurysm areas

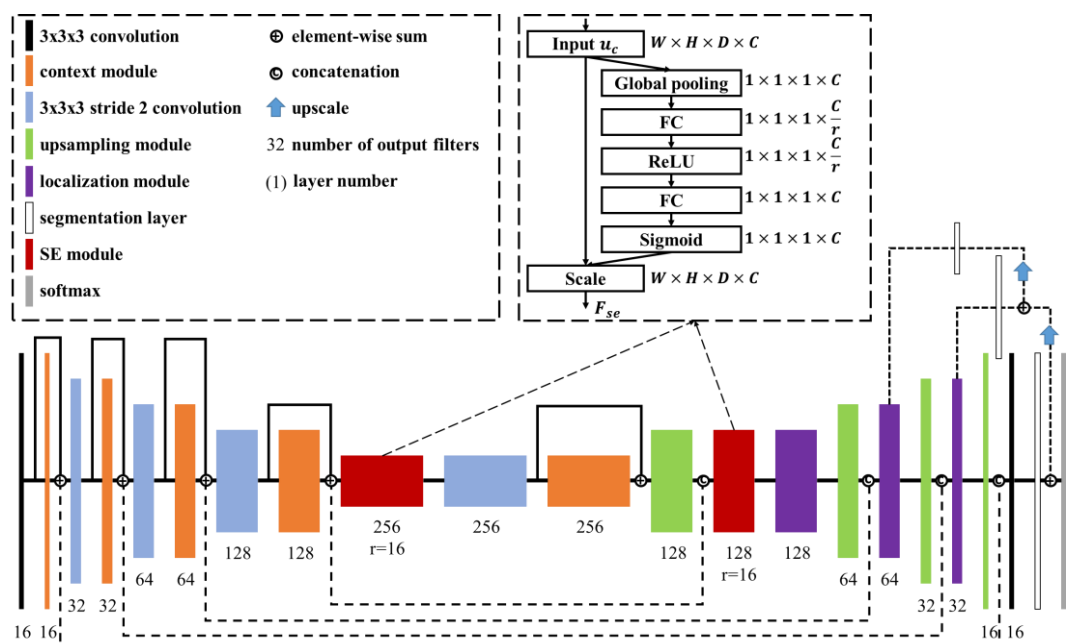

Figure 3: The proposed attention 3D U-Net network architecture

In this step, we designed an attention 3D U-Net network for the detection of aneurysms. The basic framework of the network used the improved 3D U-Net[18] proposed by Isensee et al. It first used a down-sampling path to perform high-latitude abstract encoding on the input, and then used an up-sampling path to map features that were more relevant to the task to a target of the same size as the input, thereby achieving end-to-end prediction. Downsampling was performed 4 times in the network, and the context module was used as the activation function from the input layer to collect global features and brought to the next layer. The downsampling used 3×3×3 convolution with stride 2, and the context module was composed of two 3×3×3 convolutional layers and a dropout layer (pdrop=0.3) between them. In the up-sampling process with the same number of down-sampling, the output of each up-sampling was cascaded with the output information of the context module in the down-sampling process of the corresponding level, and then output to the localization module. So that contextual information with low spatial resolution (ie high feature dimension) at a high spatial resolution could be transmitted to subsequent layers. Upsampling was to repeat the feature voxels twice, and then perform 3×3×3 convolution. The localization module was composed of 3×3×3 convolution and 1×1×1 convolution. In the upsampling process, the segmentation layers of the last three feature dimensions were combined in the form of element-wise summation to form the final network output. Moreover, all convolution calculations in the network adopt leaky ReLU nonlinearities with a negative slope of 10e-2, and used instance normalization for all batches.

Inspired by SENet[15, 22], we embedded the three-dimensional Squeeze-and-Excitation(SE) module before the last layer of the above network downsampling and after the first layer of upsampling. The module was composed of a maximum pooling layer, two fully connected layers and a ReLU layer sandwiched in between, as well as a Sigmoid layer and a Scale layer. This module first converted the high-dimensional feature maps output by the downsampling and upsampling modules into a real number sequence of 1×1×1×C (C was the feature dimension) through maximum pooling. Then used a fully connected layer with a scaling factor of R and ReLU to perform feature parameter compression and increase nonlinearity, and then connected a fully connected layer to restore the dimension. Finally, the weight of each channel was obtained through sigmoid, and the weight was added to the original feature map through the Scale operation to realize the recalibration of the original feature map.

To train our model and perform the 5-fold cross-validation, at each fold, we randomly selected 132 patients as the training set, and 34 patients as the internal test set. We augmented the 132 image sets to 1056, using flipping(by transverse section), histogram normalization, discrete Gaussian noise filter(variance: 4.0, max kernel width: 32 pixels) sequentially. To process all the labels, first calculated the center of the label area, drew a spherical area with a radius of 30 voxels from the center, and used the union of this area and the original label area as the label for actual training. The remaining blood vessel areas were all marked with another value. The image sets were then cropped and resampled to 128×128×128. Then the training sets were put into the network for training. The initial factors were: batch size = 1, initial learning rate = 5e-4, optimization function was Adam, the weights were initialized using the default initializer(glorot_uniform) of Keras. After about 244 epochs the learning process got an early stop, it cost 48.8 hours in our environment. Our environment was CPU: Intel Core i9-9900K, RAM: 32GB, GPU: NVIDIA GeForce RTX 2080Ti, Win10 professional, Tensorflow 2.0.0, Keras 2.3.1.

After training, we got the network model. And to detect the aneurysms in the TOF-MRA images, we first extracted the VOI from images, then used the model to predict each voxel left. The model would give the likelihood of each voxel to be aneurysm. Binarized the likelihood at a threshold of 0.5, a value greater than 0.5 was converted to 1 and a value less than 0.5 was converted to 0. Then, the output label image of 128×128×128 was remapped to the image of the same size as the original image according to the parameters of the previous cropping and downsampling. For the area predicted to be an aneurysm, took the center of the area and calculate the maximum distance from the center to the boundary of the area. If the distance was not greater than 30 voxels in transverse section, drew a cube area with size 60×60×H as the prediction result (length×width×height, length and width were the size in transverse section, and height refers to how many transverse-sectional slices the predicted area contains), else, the length and width were the same with the double of distance. Therefore, the cube predicted area obtained by the final processing was the detected area that might contain an aneurysm.

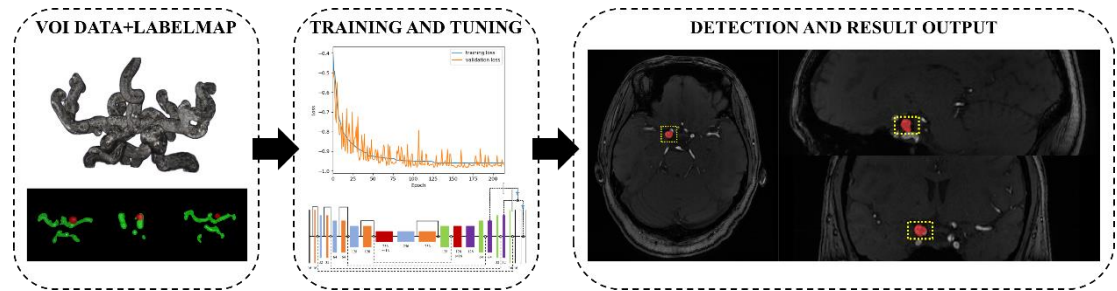

Figure 4: Workflow of the detection (in the third step, red is the output of model and yellow cube is the final output of the method)

## 3. Results

We used sensitivity and false positive rates as indicators to evaluate the proposed method, which defined as below:

$$\text{Sensitivity} = \frac{TP}{TP + FN}$$

$$\text{False positive rates} = \frac{FP}{No.\ of\ cases}$$

All the aneurysms were considered positive. As the result of the system was a cube area, if more than 30% of the aneurysm was in this cube area, then this cube area was considered a TP(True Positive) case, otherwise FP(False Positive) case.

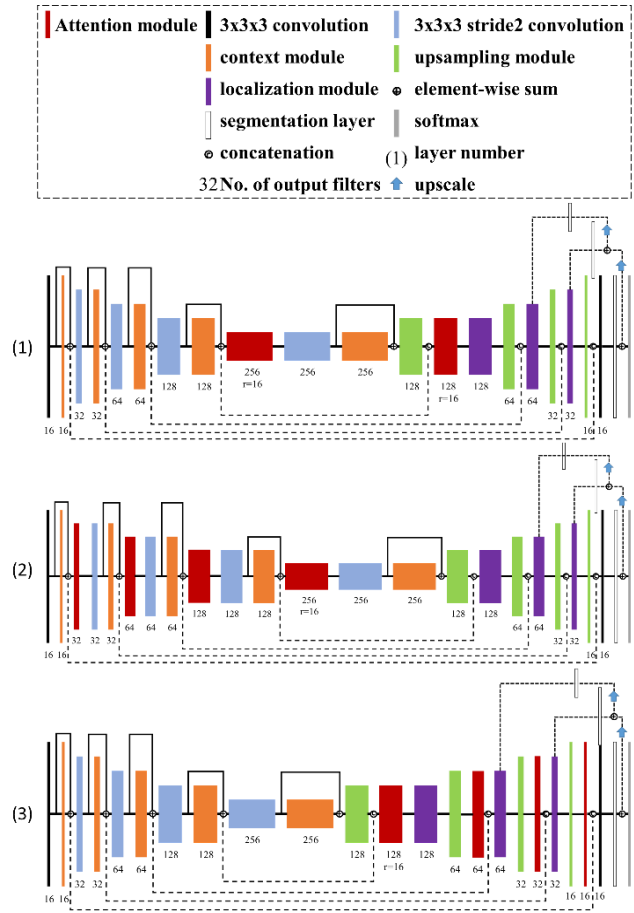

Figure 5 The three positions of the embedded attention module tested in this article: (1)Middle: in the up-down sampling conversion position; (2)Downsample: in the down-sampling path; (3)Upsample: in the up-sampling path.

In the process of selecting the attention model, we tested three attention modules of SENet[15], SKNet[14], and GCNet[16]. Among them, SENet is based on the channel attention mechanism, SKNet is based on the convolution kernel attention mechanism, and GCNet combines the Non-local module with SENet. We tried to embed GCNet, SKNet, and SENet three attention modules in the baseline network respectively, and adopted three position deployment schemes: up-sampling path, down-sampling path, and up-down sampling conversion position. The performance of the baseline network after embedding the above three attention modules at different positions was tested. In order to improve the efficiency of selection and comparison, we used the training set data of the previous paper[17], and randomly selected 38 cases in the test set of this article as the external test set for testing.

Table 3: Test results of adding different parameters of GCNet, SENet, and SKNet to different positions (Test result of the baseline network: 61.54%)

|  | Downsample | Middle | Upsample |
|---|---|---|---|
| **GCNet(R=8)** | 65.79% | 65.79% | 73.68% |
| **GCNet(R=16)** | 57.89% | 71.05% | 60.53% |
| **SKNet(R=8)** | 63.16% | 68.42% | 71.05% |
| **SKNet(R=16)** | 71.05% | 65.79% | 52.63% |
| **SENet(R=8)** | 57.89% | 73.68% | 39.47% |
| **SENet(R=16)** | 42.11% | **81.58%** | 65.79% |

According to the test results, the network performed best when the SENet module with Ratio=16 was embedded at the up-down sampling conversion position. We chose this structure as the network structure of this article, and used the training set and internal test set of this article to train the network and perform 5-fold cross-validation. In 5-fold cross-validation, the average sensitivity was 97.89±0.88%. Then we selected the model with the highest sensitivity (98.12%) in 5-fold cross-validation and tested it in the external test dataset, which was not used in the training of the model. Our method detected 61 of all the annotated aneurysms (sensitivity: 91.0%), with 2.48 false positives/case.

The 6 undetected aneurysms belong to 6 cases, among which 3 were female and 3 were male, age ranged from 42-68, the distribution covered Internal carotid artery area, Middle cerebral artery area, Anterior cerebral artery area, Posterior cerebral artery area, and the max diameter ranged from 2.67mm to 10.70mm.

In order to compare the impact of the improvement of the verification method and the increase of the data set on the performance of the method, we used the data set of the previous paper and the data set of this paper to test the method of the previous paper and the method of this paper at the same time.

Table 4: Comparison with our previous method on previous dataset and current dataset

|  | Previous Dataset[17] | Current Dataset |
| --- | --- | --- |
| **Previous Method[17]** | 82.9% | 77.6% |
| **The Proposed Method** | **88.6%(+5.7%)** | **91.0%(+13.4%)** |

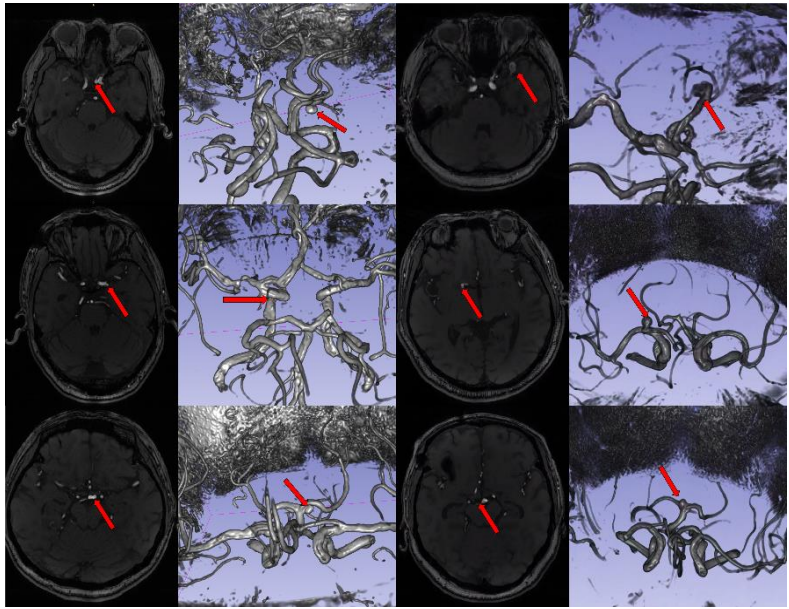

Figure 6: The undetected aneurysms in this study

## 4. Discussion and Conclusion

In this study, we designed an improved method flow for aneurysm detection, and studied the existing 3D U-Net and attention mechanism combination method, and finally improved the blood vessel extraction method, and embedded the SENet module into the baseline network. The data used in the study came from two 3.0T magnetic resonance equipment of different brands and models, and the acquisition factors of the TOF-MRA images used were also slightly different. After training the network and performing 5-fold cross-validation, the average sensitivity was 97.89±0.88%. The best model was selected and tested on the external test set. The sensitivity was 91.0% and the false positive rate was 2.48 FPs/case.

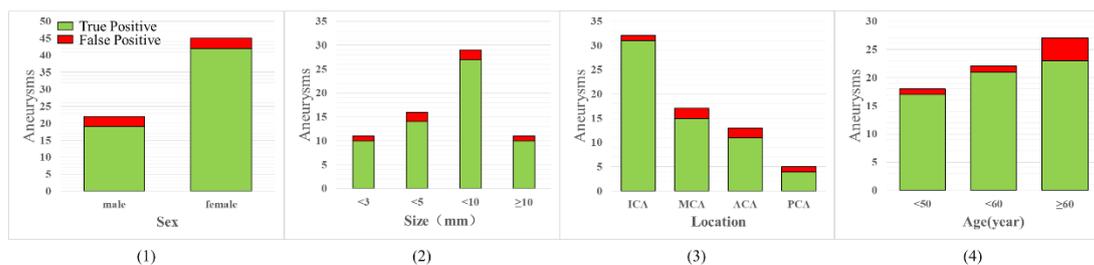

Figure 7: Subgroup analysis of sensitivity

By analyzing the result of the external test dataset, our method did not show significant differences in different sizes, different positions, and different gender subgroups. But the detection performance in cases over 60 years old is slightly worse than other age subgroups. Compared with our previous work, the method proposed this time has improvements in sensitivity, generalization, and balance. Excluding the influence of the number of cases in the data set, the method proposed this time still shows better performance. Prove the role of attention mechanism in the task of aneurysm detection. Comparing with the research results of other researchers, it can be found that this method achieves a higher sensitivity while ensuring a lower false positive rate. Since a patient usually does not have more than 3 aneurysms, when the number of false positives exceeds 3 in the test result of a case, it may cause trouble to the doctor who diagnosed it.

Table 5: Comparison with the results of other methods

|  | Sensitivity | False Positives/case | Total Cases |
|---|---|---|---|
| **Nakao[8]** | 94.2% | 2.90 | 450 |
| **Ueda[9]** | 91.0% | 6.60 | 1271 |
| **Hanaoka[10]** | 80.0% | 3.00 | 300 |
| **Faron[12]** | 90.0% | 6.10 | 85 |
| **Proposed** | 91.0% | 2.48 | 231 |

By analyzing the false-negative cases in this study, it can be found that all false-negative aneurysms are either smaller in diameter and located in the vascular bifurcation area, or larger in diameter, but with slow flow rate and density values closer to the background. Based on this, it is guessed that the network model constructed in this method has the potential to improve the recognition performance of features such as vessel length and density changes. The above problems can be solved by increasing the length measurement of similar voxels in multiple directions during the down-sampling process, averaging the density value of the extraction result of the blood vessel region, and multi-channel input. Given that the performance improvement brought by only increasing the amount of training data is not significant, it can be considered that the effective reduction of network parameters by the channel attention module is the main reason for this improvement. Therefore, for the research content of this article, effective simplification of network parameters is still the primary research direction, and the network can be improved from reducing the number of network layers and optimizing the expression of global features in the network. We will continue to conduct in-depth research on the above methods and hope to get better results.

## 5. Abbreviations

**TOF-MRA:** Time-Of-Flight Magnetic Resonance Angiography; **DSA:** Digital Subtraction Angiography; **CNN:** Convolutional Neural Networks; **FCN:** Fully Convolutional Network; **CAD:** Computer Assisted Detection; **CPU:** Central Processing Unit; **RAM:** Random Access Memory; **GPU:** Graphic Processing Unit; **MCA:** Middle Cerebral Artery; **PCA:** Posterior Cerebral Artery; **ICA:** Internal Carotid Artery; **ACA:** Anterior Cerebral Artery; **BA:** Basilar Artery; **VA:** Vertebral Artery; **FP**: False Positive Cases; **DICOM:** Digital Imaging and Communications in Medicine (DICOM) is the standard for the communication and management of medical imaging information and related data.

## 6. Acknowledge

None.

# 7. Declarations

**7.1 Ethics approval and consent to participate**

The ethics board of Huashan Hospital comprehensively re-viewed and approved the protocol of this study.

**7.2 Consent for publication**

Not applicable

**7.3 Availability of data and materials**

Not applicable.

**7.4 Competing interests**

The authors declare that they have no competing interests.

**7.5 Funding**

This work was supported by National Key Research and Development Plan [2018YFA0703101], National Natural Science Foundation of China [81971685], Shanghai Hospital Development Center [SHDC2020CR3020A], Science and Technology Commission of Shanghai Municipality [20511101100,20S31904300,19411951200], Jiangsu Key Technology Research Development Program[BE2020625,BE2018610], Suzhou Science and Technology Development Project [SS202072,SS202054,SS201854], Suzhou Health Science & Technology Project[GWZX201904], Lishui Key Technology Research Development Program [2019ZDYF09,2019ZDYF17], "Clinical Medicine Research Pilot Project" of Shanghai Medical College of Fudan University [DGF501022/015]

**7.6 Authors' contributions**

GC suggested the CAD method for cerebral aneurysm. GC, CM and XW implemented it and analyzed the images. DR, WD acquired and annotated the MR Angiography images. LY and GD reviewed the results of the imaging diagnosis. GC and YL were major contributors in writing the manuscript. All authors read and approved the final manuscript.

**7.7 Corresponding author**

Correspondence to Li Yuxin and Geng Daoying.

20 / 20

...